\documentclass[aps,twocolumn,
pre,amsmath,amssymb,showpacs,notitlepage,superscriptaddress]{revtex4-1}

\usepackage[dvipdfmx]{graphicx}

\usepackage{dcolumn}
\usepackage{bm}
\usepackage{wrapfig}
\usepackage{amsmath,amsfonts,amssymb,bm}
\usepackage[usenames]{color}
\usepackage{bbold}
\usepackage{hyperref}
\hypersetup{
    colorlinks,
    citecolor=blue,
    filecolor=blue,
    linkcolor=blue,
    urlcolor=blue
}

\newcommand{\T}{{\mathbb{T}}}
\renewcommand{\th}{{\theta}}
\newcommand{\Th}{{\Theta}}

\begin{document}

\title{Solitons in one-dimensional mechanical linkage}

\author{Koji Sato}
\email{koji.sato@imr.tohoku.ac.jp}
\affiliation{Institute for Materials Research, Tohoku University, Sendai 980-8577, Japan}
\author{Ryokichi Tanaka}
\email{ rtanaka@m.tohoku.ac.jp}
\affiliation{Mathematical Institute, Tohoku University, Sendai, 980-8578, Japan}

\begin{abstract}
It has been observed that certain classical chains admit topologically protected zero-energy modes that are localized on the boundaries. 
The static features of such localized modes are captured by linearized equations of motion, 
but the dynamical features are governed by its nonlinearity.
We study quasi-periodic solutions of nonlinear equations of motion of one-dimensional classical chains. Such quasi-periodic solutions correspond to periodic trajectories in the configuration space of the discrete systems, which allows us to define solitons without relying on a continuum theory. Furthermore, we study the dynamics of solitons in inhomogeneous systems by connecting two chains with distinct parameter sets, where transmission or reflection of solitons occurs at the boundary of the two chains. 
\end{abstract}

\maketitle

\section{Introduction}

Intriguing connections between topology and phases of matter have led to the discovery of topological insulators and superconductors~\cite{hasanRMP10,qiRMP11}.
Most of researches in this context were targeted on quantum systems, where the origin of topology comes from energy band structures. It turns out that this type of topological phenomena occur not only in quantum systems but also in classical systems, and the horizon of applications of topology has been further expanded. 

It was first shown in Ref.~\cite{kaneNatPhys13} that a certain classical system of elastic chain could support localized zero-frequency ``floppy modes", whose existence is associated to a topological index derived from the linearized equations of motion. This topological index of the elastic chain is analogous to the Su-Schrieffer-Heeger (SSH) model~\cite{suPRL79,heegerRMP88}, and it is related to the static feature of a localized mode, such as where it is localized~\cite{kaneNatPhys13}. In contrast to the static behavior which can be explained by the linearized equations motion, the description of its dynamical feature requires full nonlinear aspect of the equations~\cite{chenPNAS14}. In Ref.~\cite{chenPNAS14}, they derived the $\phi^4$ theory or the sine-Gordon theory via continuum limit, depending on the set of parameters in the discrete system. These nonlinear equations support kink solutions, which describe the propagation of localized modes. However, these solutions do not capture the full dynamics of the localized modes; the dynamics of the localized modes in the original discrete systems are actually periodic or quasi-periodic, but the kink solutions of the aforementioned continuum equations do not capture this periodic feature.    

In this paper, we are interested in the full behavior of quasi-periodic solutions in discrete systems. It turns out that the global dynamics of such quasi-periodic solutions are governed by the topology of the phase space of mechanical linkages (networks formed by rigid bars and joints). In fact, from this point of view, the kink solutions in the continuum limit capture only a part of the full dynamics; this is because these solutions need to take account of not only the nature of propagation but also the nature of recurrence~\footnote{It is not clear that an appropriate boundary condition in the continuum theory can be imposed to capture the nature of recurrence or the full dynamics of the original discrete system. Certain features of the full dynamics, such as scattering of kink and antikink solutions by impurities, can be analyzed by a continuum equation~\cite{zhouPRE17}.}. In order to describe the full dynamics, we formulate soliton solutions without taking the continuum limit. We further clarify the connection between solitons and the configuration space of linkages. This also enables us to deal with inhomogeneous systems which are not discussed in Ref.~\cite{chenPNAS14}.

Our understanding provides a geometrical perspective of quasi-periodic solutions. The linearized equations of motion correspond to the tangent space of the configuration space, and the soliton solutions can be viewed as quasi-periodic one-dimensional trajectories in the configuration space. Such one-dimensional trajectories are obtained by the constraint conditions imposed on the mechanical chain. The topology of the one-dimensional trajectories in the configuration space depends on the parameters of the system, and different types of soliton modes are captured by its topology~\cite{chenPNAS14}. We further consider a system consisting of two distinct chains with different parameter sets. As a result, we find rich properties of soliton propagation depending on how we combine two different chains.  

In fact, the model we are considering can be regarded as a variant of a classical one-dimensional chain of coupled oscillators. Such systems are described by equations of motion of the form, $\ddot x_i=V'(x_{i+1}-x_i)-V'(x_{i}-x_{i-1})$, where $x_i=x_i(t)$ ($i=1,\cdots,N$) is the position of $i$-th mass at time $t$ with nearest neighbor potentials $V(x_{i+i}-x_i)$. For example, Fermi, Pasta and Ulam~\cite{Fermi} took the potential of the form given by $V(y)=\frac{1}{2}y^2+\frac{\alpha}{3}y^3 + \frac{\beta}{4}y^4$ with $N=64$. Our system corresponds to a one-dimensional elastic chain with the nearest neighbor potentials $U(x_i, x_{i+1})$ with strong nonlinearity. The model of this type is first studied by Kane and Lubensky~\cite{kaneNatPhys13}, and subsequently by Chen {\it et al.} regarding its dynamical perspective~\cite{chenPNAS14}. We further investigate this model and systematically construct robust quasi-periodic solutions from the topological structure of the configuration space.

In Sec.~\ref{sec2}, we review the linearized description of the elastic chain considered by Kane and Lubensky in order to refer some of the results here for the later analysis. In Sec.~\ref{sec3}, we develop the idea of how to view solitons without taking a continuum limit by introducing the configuration space of the discrete linkage system. Finally, in Sec.~\ref{sec4}, we investigate the transport property of solitons in hybrid chains, which are constructed by connecting two distinct chains with different parameters. 

\section{Mechanical chains with linearized description}\label{sec2}

\subsection{Kane-Lubensky chain: linearized equations and edge modes}

One of the ways to characterize the stability of mechanical structures is to examine the relation among the spatial dimension of the system, $d$, and the number of mass points ($N_{\rm s}$) and bonds ($N_{\rm b}$). The number of zero mode, denoted by $N_0$, is related to these parameters of a given system by $N_0=dN_{\rm s}-N_{\rm b}$, known as Maxwell's rule~\cite{maxwell,calladine}. Here, $dN_{\rm s}$ is meant to be the number of degrees of freedom of a mass point under constraint conditions~\cite{paulose15}. The presence of such a zero mode is associated with mechanical instability of the system.  
This relation was modified to a more general form~\cite{kaneNatPhys13}: $\nu=N_0-N_{\rm ss}=dN_{\rm s}-N_{\rm b}$ with $\nu\in\mathbb Z$, where $N_{\rm ss}$ is the number of states of self-stress. In the presence of the states of self-stress, bonds are under tension or compression even in the absence of net force on each mass point. Here, $\nu$ can be interpreted as the topological index which is derived from the set of linearized equations of motion of the mass points, as shown below.

Given a system of a mechanical network, when the tension $T_m$ in the $m$-th bond and the force $F_i$ acting on the $i$-th site are present, they are linearly related by $F_i=Q_{im}T_m$. Under a displacement of the mass $u_i$ at the $i$-th site, the extension $e_m$ of the $m$-th bond is then expressed by $e_m=Q^T_{mi}u_i$. The null space of $Q$ corresponds to a state of the network under non-zero tension on the bonds and zero net force on every mass point. This is the state of self-stress, and the number of such states is given by $N_{\rm ss}={\rm dim}\,{\rm Ker}\,Q$. The null space of $Q^T$ corresponds to a state with finite displacements of mass points in the absence of tensions in all the bonds, which is called the zero mode. The number of zero modes is thus given by $N_0={\rm dim}\,{\rm Ker}\,Q^T$. Maxwell's rule can thus be stated by $\nu=N_0-N_{\rm ss}={\rm dim}\,{\rm Ker}\,Q^T-{\rm dim}\,{\rm Ker}\,Q$. 
Note that the difference ${\rm dim}\,{\rm Ker}\,Q^T-{\rm dim}\,{\rm Ker}\,Q$ can remain constant even under changing parameters of a given system.

\begin{figure}[htb]
\begin{center} 
\includegraphics[width=1\linewidth]{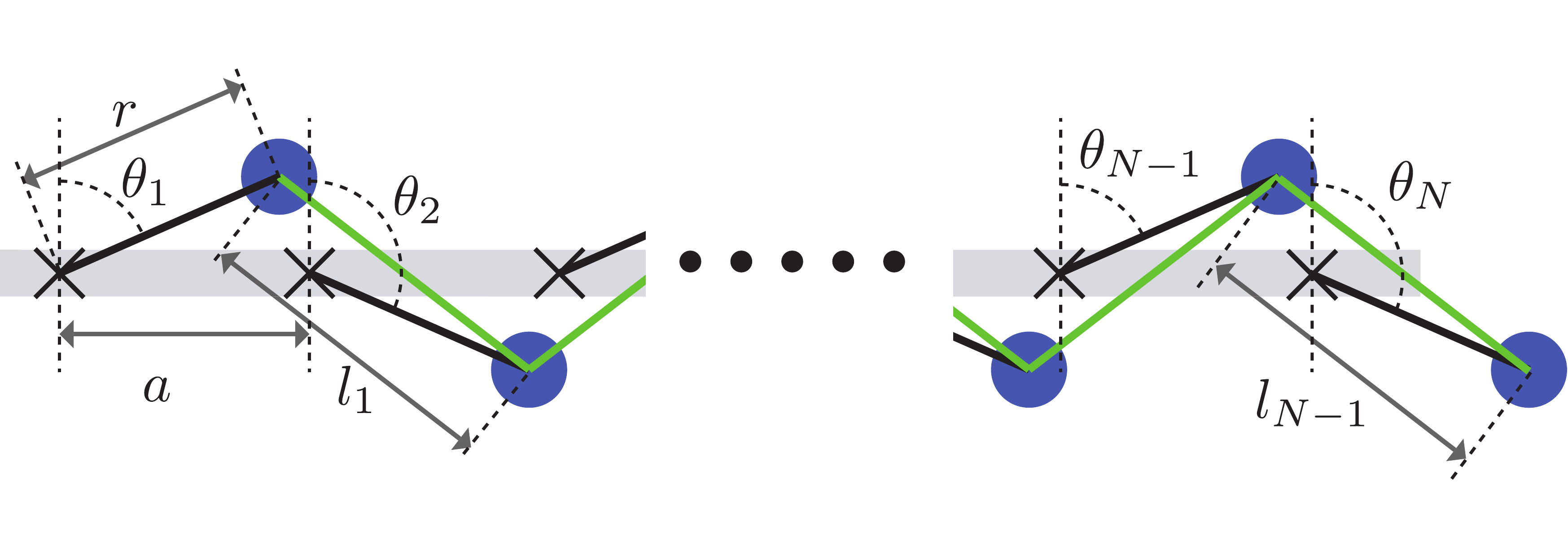}
\caption{Schematic of the linkage system. $\theta_i$ is the angle associated to the $i$-th rotor, $a$ the separation between the adjacent pivot points, $r$ the length of the arm, and $l_{i}$ the bond length between the adjacent mass points at the $i$-th and the $(i+1)$-th sites. }
\label{fig_linkage}
\end{center}
\end{figure}
In Ref.~\cite{kaneNatPhys13}, Kane and Lubensky considered a type of one-dimensional elastic system with $N$ number of rotors with equal arm length, which are placed with equal distance (see Fig.~\ref{fig_linkage}). More generally, let us denote the distance between the $i$-th  and $(i+1)$-th pivots as $a_i>0$ and the $i$-th arm length as $r_i$ in order to treat the cases of non-uniform parameters.
Each arm has a mass point with mass $M$. The $i$-th mass is allowed to rotate around the $i$-th pivot, which acts as a rotor. The angle $\theta_i$ of the $i$-th rotor is measured with respect to the vertical axis.  The bond length between the $i$-th and the $(i+1)$-th mass points is then given by $l_{i}(\theta_i,\theta_{i+1})$.
We denote a given set of initial angles of $N$ number of rotors by $\bar\theta=\{\bar\theta_1,\cdots,\bar\theta_{N}\}$, which sets the equilibrium length of the $i$-th bond by $\bar l_{i}=l_{i}(\bar\theta_i,\bar\theta_{i+1})$.
Assuming that all the bond lengths stay constant, namely $l_{i}=\bar l_{i}$ for all $i$, we deduce that infinitesimal changes of the angles $\theta_i=\bar\theta_i+\delta\theta_i$ satisfies that
\begin{eqnarray*}
0&=&l_i(\bar\theta_i+\delta\theta_i,\bar\theta_{i+1}+\delta\theta_{i+1})-\bar l_i\\
&=&p_{i}(\bar\theta)\delta\theta_{i}+q_{i}(\bar\theta)\delta\theta_{i+1}+\mathcal O(\delta\theta^2)
\end{eqnarray*}
where we defined the following functions:
\begin{eqnarray}\label{Eq:p_q}
p_i(\theta)&=&\frac{r_i}{\bar l_{i}}\left[-a_i\cos \theta_i +r_{i+1}\sin(\theta_i-\theta_{i+1})\right],\nonumber\\
q_i(\theta)&=&\frac{r_{i+1}}{\bar l_{i}}\left[a_i \cos \theta_{i+1} - r_i \sin(\theta_i-\theta_{i+1})\right].
\end{eqnarray}
Writing $(N-1)\times N$-matrix $Q^T(\theta)=(Q^T_{m i}(\theta))$, one can cast $(N-1)$ number of the constraint conditions into the following form: 
\begin{align}\label{Eq:linear}
&\sum_{i=1}^NQ^T_{m i}(\bar\theta)\delta\theta_i=0\,,\quad
\text{for $m=1,\cdots, N-1$}\,.
\end{align}  
Given a generic configuration with $a_i$, $r_i$ and $\bar\theta_i$, the rank of the matrix $Q^T(\bar\theta)$ is $N-1$, thereby the null space is one dimensional.
One may typically observe that a non-zero solution of Eq.~(\ref{Eq:linear}) is exponentially localized on one of the two ends. 
Such a solution is called an {\it edge mode} in the mechanical chain, which is analogous to topological edge modes appearing in electronic systems,  e.g. SSH model.

If the bond length is allowed to change, the infinitesimal bond extensions are given by
\begin{equation}
e_m=\sum_{i=1}^NQ^T_{mi}(\bar\theta)\delta\theta_{i}\,.
\end{equation}
Letting $K$ be the spring constant of the bonds connecting the adjacent mass points, we approximate the change in the length of $i$-th bond by
\[l_i(\theta)-\bar l_i(\bar\theta) \approx \sum_{k=1}^NQ_{i k}^T(\bar\theta)\left(\theta_k -\bar\theta_k\right)\,.\]
The corresponding Lagrangian is then given by
\begin{equation}
L=\sum_{i=1}^N\frac{1}{2}Mr_i^2\left(\frac{d\theta_i}{dt}\right)^2-\frac{1}{2}K\sum_{i=1}^{N-1}\left[Q_{i k}^T(\bar\theta)\left(\theta_k -\bar\theta_k\right)\right]^2\,,
\end{equation}
and the equations of motion of the angular variables $\theta_i(t)$ for $i=1,\cdots,N$ are obtained as
\begin{align}\label{eom}
\frac{d^2\theta_i(t)}{dt^2}&=-\frac{K}{Mr^2_i}\sum_{j=1}^N\left[Q\left(\bar\theta\right)Q^T\left(\bar\theta\right)\right]_{ij}\left[\theta_j(t)-\bar\theta_j\right]\\
&\hspace{2cm}\text{for $i=1, \dots, N$.}\nonumber
\end{align}
If $Q^T(\th)$ has rank $N-1$, then ${\rm dim}\,{\rm Ker}\,Q^T(\th)=1$ and ${\rm dim}\,{\rm Ker}\,Q(\th)=0$; this is indeed the case for generic $\th$ and in this case we have
$\nu={\rm dim}\,{\rm Ker}\,Q^T(\th)-{\rm dim}\,{\rm Ker}\,Q(\th)=1$. 
For some $\th$, the matrix $Q^T(\th)$ may have rank less than $N-1$, but  
the index remains $\nu=1$ for every $\th$.

\section{Nonlinear mechanical chains and solitons}\label{sec3}

\subsection{Spectral gaps and configuration spaces of four-bar linkages}

\begin{figure}[htb]
\begin{center} 
\includegraphics[width=.5\linewidth]{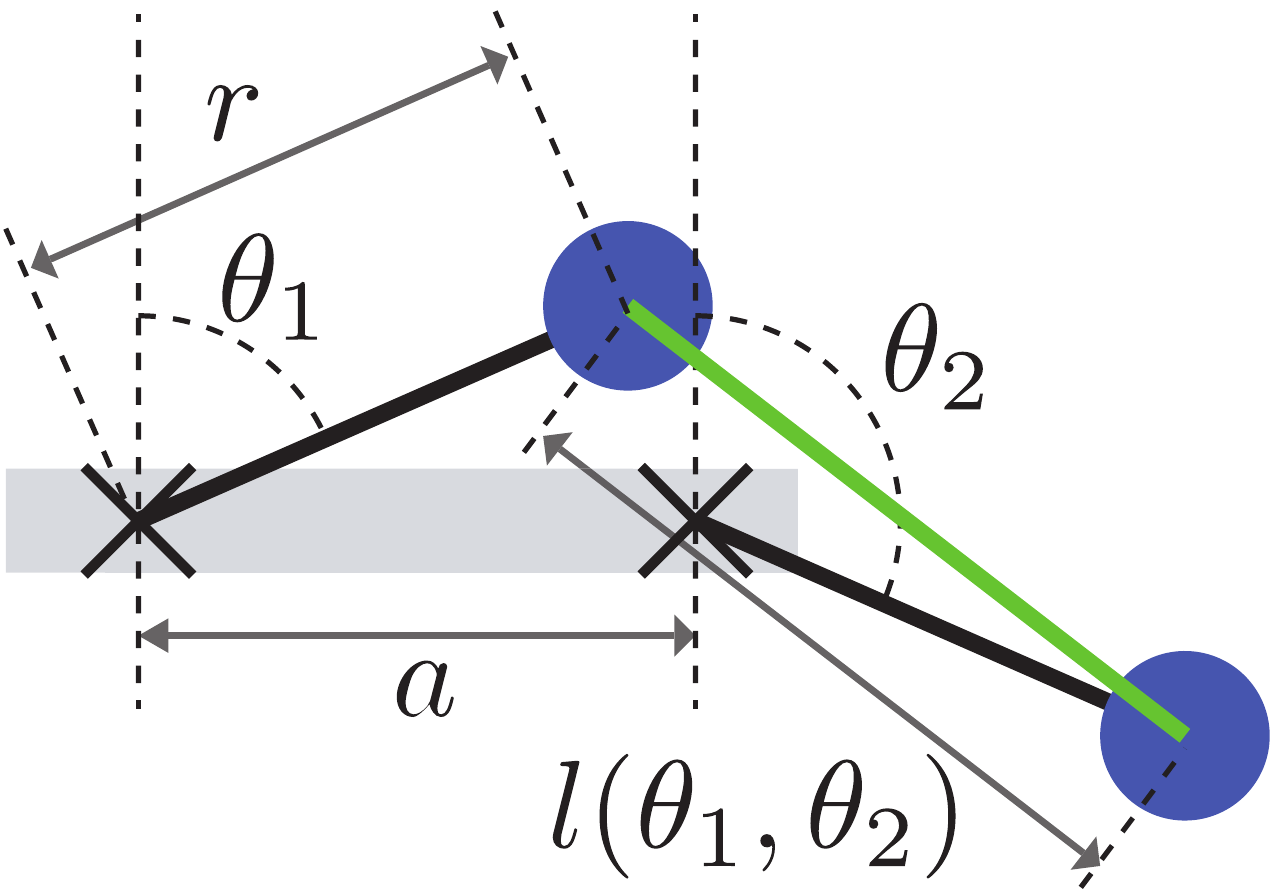}
\end{center}
\caption{A single four-bar linkage with the arm lengths given by $r$ and the separation of pivots given by $a$. The length of the bond (green) connecting two mass points (blue) is given by $l(\theta_1, \theta_2)$.}\label{fig-four-bar}
\end{figure}

First we shall analyze the special case that the parameters of the system are uniform, namely $a=a_i$, $r=r_i$ and $l_i(\bar\theta)=\bar l$ for all $i=1, \dots, N$.
In this case, all the possible configurations of this uniform linkage are determined by a single four-bar linkage with two rotors (see Fig.~\ref{fig-four-bar}).
We give a belief description on the configuration space of four-bar linkages below and  introduce a more elaborate description of the configuration space of linkages.

\begin{figure}[htb]
\begin{center}
\includegraphics[width=1\linewidth]{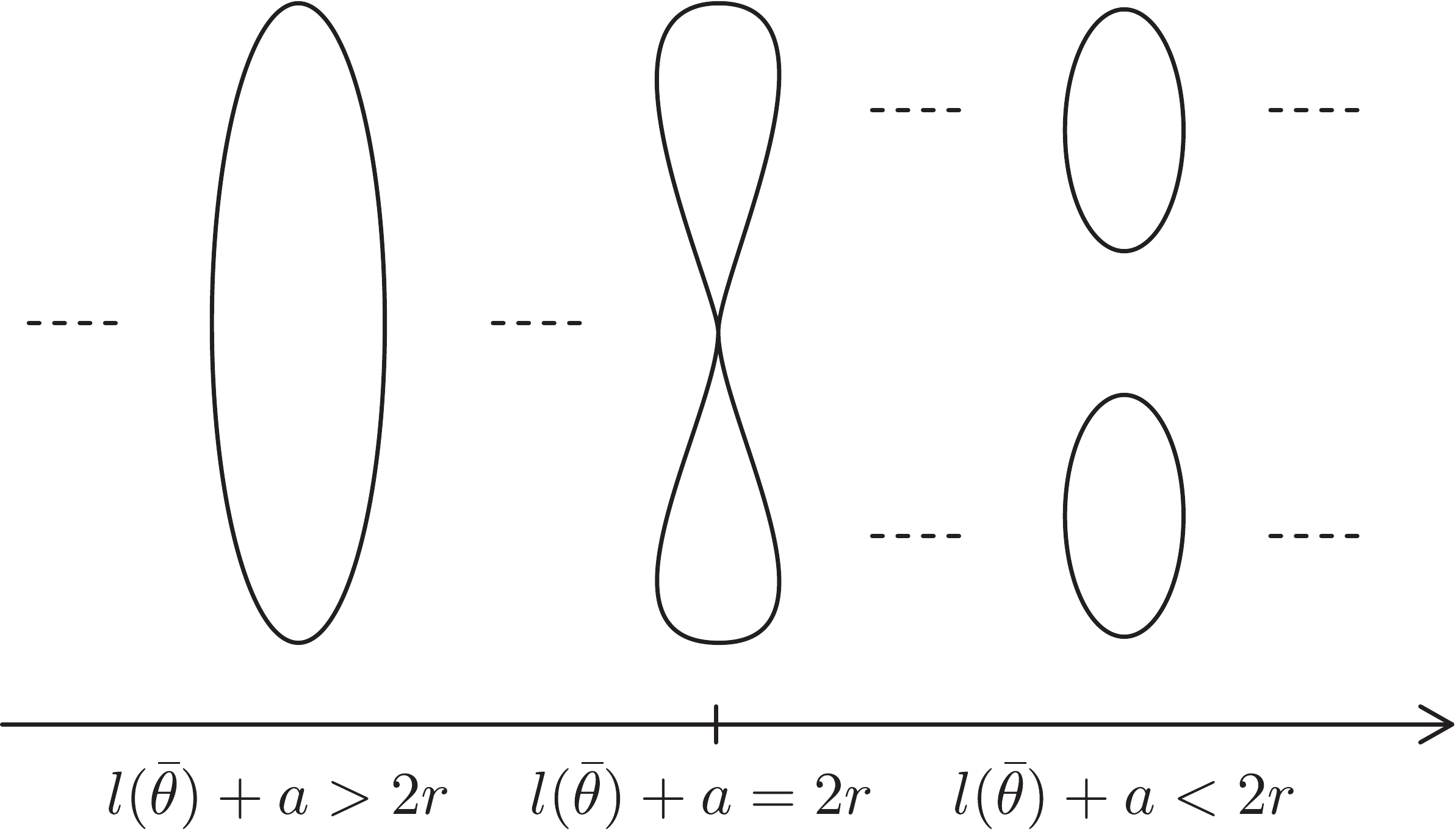}
\end{center}
\caption{The configuration spaces a four-bar linkage, and its dependence on the parameters.}\label{fig_morse}
\end{figure}

After determining a set of initial angles, denoted by $\bar\theta=(\bar\theta_1, \bar\theta_2)$, we obtain a single constraint condition for the bond length given by $l(\theta_1, \theta_2)=l(\bar\theta)$ with
\begin{eqnarray}\label{Eq:length}
&&l(\theta_1, \theta_2)\\ \nonumber
&&=\sqrt{\left[a+r\left(\sin \th_2-\sin \th_1\right)\right]^2+r^2\left(\cos \th_1 - \cos \th_2\right)^2}\,.
\end{eqnarray}
The configuration space of a four-bar linkage is the space of shapes which are continuously deformed under the constraint fixing the lengths of all four sides of the linkage. Since the constraint condition relates two angular variables, specifying a single angle determines the entire shape of the four-bar linkage. This means that the configuration space is a union of circles, and its structure depends on the lengths of $a$ and $r$.
If the four-bar linkage is close to a triangle (e.g. small $a$ and large $r$), the configuration space is disconnected and consists of two disjoint circles. This is the case of $l(\bar\theta)+a<2r$ in Fig.~\ref{fig_morse}; then a certain linkage configuration cannot be continuously deformed to its mirror image.
On the other hand, the configuration space is connected and consists of a single circle if the four-bar linkage is close to a rectangle (e.g. large $a$ and small $r$), which corresponds to the case of $l(\bar\theta)+a>2r$ in Fig.~\ref{fig_morse}.
It is a union of two circles jointed at a single point, when $l(\bar\theta)+a=2r$. In this case, there is a singularity at the joint point of the two circles.  

More precisely, in our case, the configuration space is disconnected if and only if $l(\bar\theta)+a < 2r$~\cite{KapovichMillson}.
We state that the linkage is in a {\it disconnected phase} (or {\it spinner phase}) when its configuration space is disconnected, and similarly a {\it connected phase} (or {\it flipper phase}) for a connected configuration space, following the terminology introduced in Ref.~\cite{chenPNAS14}.
The disconnected phase with $l(\bar\theta)+a < 2r$ can be restated as $a/r <\sin^2 \bar\theta$, and a connected phase with $l(\bar\theta)+a > 2r$ leads to $a/r>\sin^2\bar\theta$. When $a/r=\sin^2 \bar\theta$, the configuration space is a union of two circles at a single point, which represents the critical phase separating two topologically distinct structures of the configuration space. 
In summary, we have
\begin{equation}\label{phase_condition}
\begin{cases}
&\text{disconnected (spinner) phase}\\
&\hspace{1cm}\Leftrightarrow \quad l(\bar\theta)+a < 2r  \quad \Leftrightarrow\quad  1 <\frac{r}{a}\sin^2 \bar\theta\,, \\
&\text{connected (flipper) phase}\\
&\hspace{1cm}\Leftrightarrow\quad  l(\bar\theta)+a > 2r  \quad \Leftrightarrow\quad  1 > \frac{r}{a}\sin^2 \bar\theta\,. \\
\end{cases}
\end{equation}

Under a given set of initial conditions of angles and angular velocities, the shape of a four-bar linkage evolves within the configuration space. There is a further characterization of this dynamics based on the relative sign of initial angular velocities of the two rotors~\cite{chenPNAS14}. When a four-bar linkage has a disconnected configuration space, the initial angular velocities of the two rotors always have the same sign, which we call {\it non-repulsive condition} ({\it wobbling condition} in \cite{chenPNAS14}). When a four-bar linkage has the connected configuration space, there are configurations that two rotors can admit initial velocities with opposite signs, and such a case is called {\it repulsive condition} ({\it non-wobbling condition} in \cite{chenPNAS14}). 
In the case of uniform linkage structure where $a_i=a$ and $r_i=r$ for all $i$ under a given set of initial angles, one may find $p_i(\bar\theta)$ and $q_i(\bar\theta)$, and in turn also determine $Q^T(\bar\theta)$. From the condition in Eq.~(\ref{Eq:linear}), the initial angular velocities of the rotors are in the null space of $Q^T(\bar\theta)$. We can then find that the initial system satisfies the repulsive (non-repulsive) condition where the initial angular velocities of a give adjacent rotors have the opposite (same) signs when $p_i(\bar\theta)/q_i(\bar\theta)<0$ ($p_i(\bar\theta)/q_i(\bar\theta)>0$). This condition can be also restated as $a<2r\sin\bar\theta$ ($a>2r\sin\bar\theta$) for the repulsive (non-repulsive) condition~\cite{chenPNAS14}, which is summarized as
\begin{equation}\label{wobbling_condition}
\begin{cases}
&\text{repulsive (non-wobbling) condition}\  \Leftrightarrow\  1<\frac{2r}{a}\sin\bar\theta\,, \\
&\text{non-repulsive (wobbling) condition}\  \Leftrightarrow\  1>\frac{2r}{a}\sin\bar\theta\,. \\
\end{cases}
\end{equation}

The topology of the configuration space is also captured by the spectrum of the dynamical matrix given by $Q(\th)Q^T(\th)$.
In the case of a four-bar linkage $(N=2)$, the matrix $Q(\th)Q^T(\th)$ is a square matrix of dimension $2$.
The spectrum is given by $0$ and $\lambda_1(\th)$ where
\begin{equation}\label{Eq:spec}
\lambda_1(\th)=\left(\frac{\partial l}{\partial \th_1}(\th)\right)^2+\left(\frac{\partial l}{\partial \th_2}(\th)\right)^2.
\end{equation}
Note that $\lambda_1(\th)>0$ for all $\th$ if and only if $\frac{\partial l}{\partial \th_1}(\th) \neq 0$ or $\frac{\partial l}{\partial \th_2}(\th) \neq 0$ for all $\th$. 
If $\lambda_1(\th)=0$ for some $\th$ if and only if $\frac{\partial l}{\partial \th_1}(\th) = \frac{\partial l}{\partial \th_2}(\th) = 0$ for some $\th$, in which case, 
$\th=\pm \frac{\pi}{2}$
and
$l(\th)=|2r-a|$, i.e.\ the critical case.

In a more general case of a chain with $N$ number of rotors, the first non-zero eigenvalue $\Lambda_1(\th)$ (counted with multiplicity) of $Q(\th)Q^T(\th)$ is positive $\Lambda_1(\th)>0$ for $\th=(\th_1, \dots, \th_N)$ if and only if ${\rm rank}\,Q^T(\th)=N-1$. In particular, if $\Lambda_1(\th)>0$ for all $\th$, then the configuration space of the corresponding linkage has no singularity.

\subsection{A nonlinear potential and quasi-periodic solutions}

\begin{figure*}[htb]
\begin{center} 
\includegraphics[width=1\linewidth]{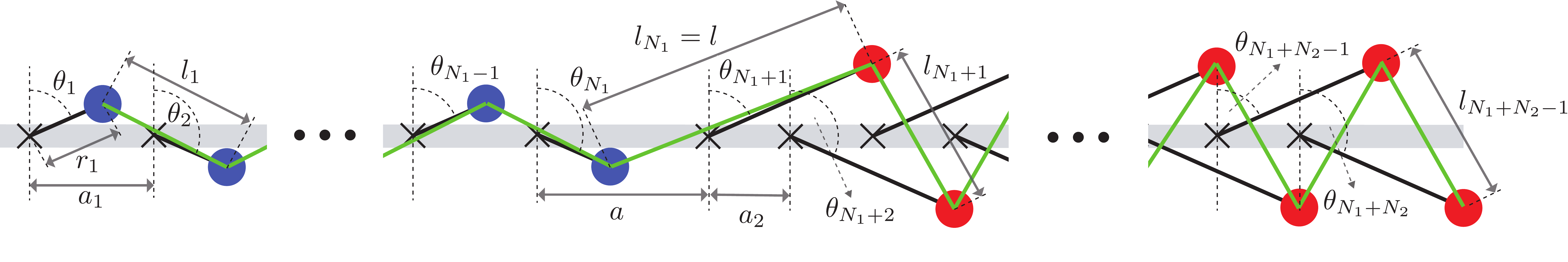}
\caption{Two sub-linkages with different parameter sets are connected to make a hybrid linkage system. The above figure illustrates the case that the left (right) sub-linkage, marked in blue (red), is in the connected (disconnected) phase. $r_{1(2)}$ and $l_{1(2)}$ are the arm length and equilibrium bond length for the left (right) sub-linkage. $\theta_i$ is the angle of the rotor at the $i$-th site measured with respect to the vertical line, and $i$-th bond length is denoted by $l_i$. $l$ is the length of the bond connecting the left (blue) and right (red) sub-linkages.}\label{fig2}
\end{center}
\end{figure*}

In order to treat the dynamics of the linkage, we approximate the constraint condition of constant bond length and then define the potential term following Ref.~\cite{chenPNAS14}. We suppose that the bond lengths are almost unchanged:
\begin{equation}\label{Eq:potential}
\frac{1}{2}K\left[l_{i}(\theta_i,\theta_{i+1})-{\bar l}_{i}\right]^2 \approx \frac{K}{8{\bar l}_{i}^{\ 2}}\left[l_{i}^2(\theta_i,\theta_{i+1})-{\bar l}_{i}^{\ 2}\right]^2,
\end{equation}
so that the corresponding Lagrangian has the form
\begin{equation}\label{modified_lagrangian}
L=\frac{1}{2}\sum_{i=1}^N M r_i^2\left(\frac{d\theta_i}{dt}\right)^2-\sum_{i=1}^{N-1}\kappa_i\left[l_{i}^2(\theta_i,\theta_{i+1})-\bar l_{i}^{\ 2}\right]^2,
\end{equation}
where $\kappa_i=K/8{\bar l}_{i}^{\ 2}$. The dynamics generated by the resulting equations of motion are not exactly what is given by strictly following the constraint conditions $l_i(\theta_i,\theta_{i+1})=\bar\theta_i$ for all $i$, but its trajectory provides a good approximation for the exact one. The initial velocities of the $N$ rotors are taken from ${\rm Ker\,}Q^T(\theta)$, so that Eq.~(\ref{Eq:linear}) is satisfied. Then the associated dynamics will stay close to the trajectory given by the constraints $l_{i}(\theta_i,\theta_{i+1})={\bar l}_{i}$ for all $i$. This is because of the stability of the solution, which is guaranteed by the fact that ${\rm Ker\,}Q^T(\theta)={\rm Ker\,}Q(\theta)Q^T(\theta)$ and all the other eigenvalues of $Q(\theta)Q^T(\theta)$ are positive.  One can then observe a quasi-periodic motion of the linkage. It is generally impossible to obtain an analytic form of the solution, but the trajectory of this dynamics can be understood through the topology of the configuration space on the four-bar linkage, which we describe below.

The coordinate system of total angles $\theta=(\theta_1, \dots, \theta_N)$ is given by the $N$-dimensional torus $\T^N$.
For each parameter set of the system $\Theta=\{a_i, r_i, \bar\theta_i\}_{i=1, \dots, N}$,
the space of all configurations of the entire linkage system is the subspace $C(\Theta)$ defined by the set of constraints
$l_{i}(\th_i, \th_{i+1})=\bar l_{i}$ for all $i$.
Although the space of coordinates is $N$-dimensional, the subspace $C(\Theta)$, obtained under the constrains, is a one-dimensional subspace,
along which the dynamics of the linkage propagates.
Depending on the parameter set $\Theta$, the configuration space $C(\Theta)$ has a topologically distinct realization, and this affects the dynamics of the trajectory.
In the case of four-bar linkage, the configuration space $C(\Th)$ is topologically either a circle or the disjoint union of two circles depending on the parameter set $\Th$ as in (\ref{phase_condition}).

Let us denote the trajectory parametrized by time $t$ by $\th(t)=(\th_1(t), \dots, \th_N(t))$ in $C(\Th)$.
Given the $(N-1)\times N$-matrix $Q^T(\th)$ in Eq.~(\ref{Eq:linear}), the angular velocity vector $\dot{\th}(t)=(\dot{\th}_1(t), \dots, \dot{\th}_N(t))$ is in the null space of $Q^T(\th)$. The trajectory $\th(t)$ satisfies the equation
\begin{align}\label{Eq:soliton}
&\sum_{i=1}^NQ^T_{m i}\left(\th(t)\right)\dot{\th}_i(t) =0\,, \quad\text{for $m=1,\cdots,N-1$}\,.
\end{align}
We say that a trajectory $(\th(t))_{t \in [0, T]}$ is {\it generic} if the matrix $Q^T(\th(t))$ depends on $t$ without singularities and the rank of $Q^T(\th(t))$ is $N-1$ for all $t$,
and call a nontrivial generic trajectory a {\it soliton}.

The properties of solitons in a linkage were extensively analyzed in Ref.~\cite{chenPNAS14}. Since such modes of solitons have rich properties, which depend on sets of parameters, one may wonder how solitons would behave if a ``domain wall" is introduced by connecting two linkage systems (with the same or different parameter sets) by a bond with an arbitrary length.  For instance, we can consider connecting two distinct linkages in disconnected and connected phases as shown in Fig.~\ref{fig2}.

\section{Theoretical analysis and simulation of solitons in hybrid systems}\label{sec4}
\begin{figure*}[htb]
\begin{center} 
\includegraphics[width=1\linewidth]{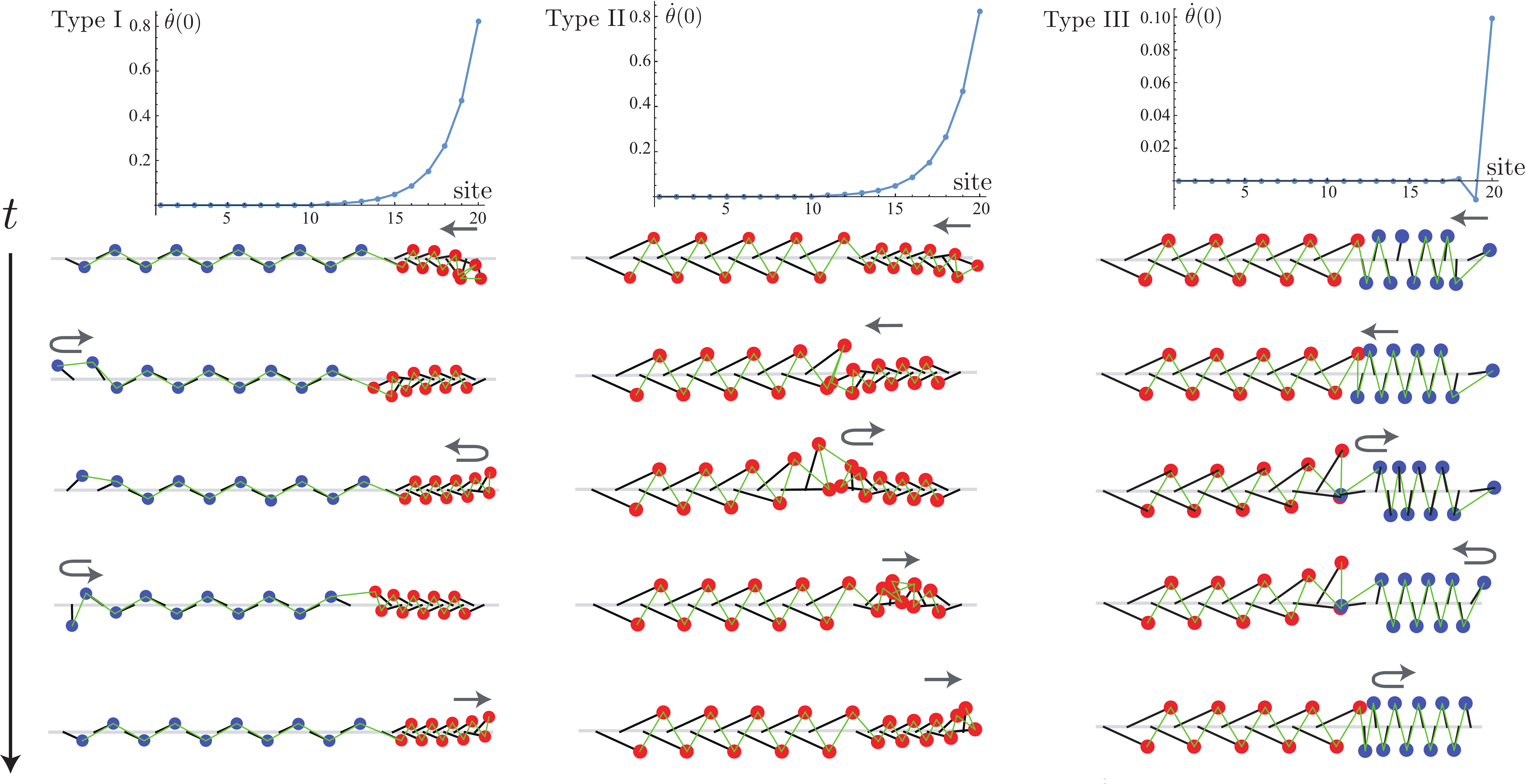}
\caption{In Type I, we have combined a connected/non-repulsive phase ($a_1=1.5$, $r_1=1$, $\bar\theta_i=2$ for $i=1, \dots, 10$) and a disconnected/repulsive phase ($a_2=0.5$, $r_2=1$ and $\bar\theta_i=2$ for $i=1, \dots, 10$) with spacing $a=2$.
In Type II, we have combined two disconnected/non-repulsive phases ($a_1=1$, $r_1=2$, $\bar\theta_i=2$ for $i=1, \dots, 10$) and ($a_2=0.5$, $r_2=1$ and $\bar\theta_i=2$ for $i=1, \dots, 10$) with spacing $a=2$. In Type III, we have combined a disconnected/non-repulsive phase ($a_1=1$, $r_1=2$, $\bar\theta_i=2$ for $i=1,\cdots,10$) and a connected/repulsive phase ($a_2=0.5$, $r_2=1$, $\bar\theta_i=0.2$ for $i=1,\cdots,10$) with spacing $a=2$. 
In all the cases, we took the mass and spring constant as $M=0.01$ and $K=10$, respectively. The graphs on the top show the profile of the initial angular velocities, $\dot\theta(0)$ versus site locations, given by Eq.~(\ref{Eq:soliton}) with the initial angles $\bar\theta_i$ for $i=1,\cdots,20$ shown above for each case. The figures of the linkages show the time evolution of a soliton in each case. 
The numerical simulation is taken with time discretization $\Delta t=0.01$ and number of steps $1.0\times 10^4$.
}\label{fig_soliton_types}
\end{center}
\end{figure*}
\begin{table*}[htb]
\begin{tabular}{|l||c|c|c||c|c|c|}
 \hline 
\rule[-0.3cm]{0cm}{0.8cm}
		&	Left phase & $\dfrac{r}{a}\sin^2 \bar\theta$ & $\dfrac{2r}{a}\sin \bar\theta$	&	Right phase & $\dfrac{r}{a}\sin^2 \bar\theta$ &$\dfrac{2r}{a}\sin\bar\theta$ \\ \hline \hline
Type I	&	Connected/Non-repulsive		& 0.551 & 1.212 &	Disconnected/Non-repulsive	& 1.653	 & 3.637\\ \hline
Type II	&	Disconnected/Non-repulsive		& 1.654 & 3.637 & 	Disconnected/Non-repulsive	& 1.654 & 3.637	\\ \hline
Type III	&	Disconnected/Non-repulsive			& 1.654 & 3.637 & 	Connected/Repulsive	&  0.079 & 0.795\\ \hline
\end{tabular}
\caption{Summary of the information the left and right linkages of hybrid linkage systems corresponding to Type I, II, and III. The numbers in the table are to be referred to the conditions (\ref{phase_condition}) and (\ref{wobbling_condition}) to determine the phase of left or right part of a given hybrid linkage.}\label{phase_table}
\end{table*}

We now consider a hybrid linkage structure by combining two sub-linkage systems, denoted by sets of parameters $\Theta_1$ and $\Theta_2$, and connect them by a bond. The subsystem $\Theta_k$ ($k=1,2$) consists of $N_k$ number of the pivots with arms of equal length $r_{k}$ are placed with equal spacing $a_{k}$ with initial angles $\bar\theta_k$.
Two systems $\Theta_1$ and $\Theta_2$ are placed with the middle spacing $a$ and glued by the bond with the length, $l$, determined by the initial condition.
This combined system $\Theta$ denoted by $(\Theta_1, \Theta_2, a)$ will admit a soliton solution if Eq.~(\ref{Eq:linear}) has an edge mode and there is a generic solution for Eq.~(\ref{Eq:soliton}).
There appear to be a number of different modes of propagation of solitons in this hybrid linkage, and complete classification of these solitons is beyond the scope of this work. In order to discuss the structure of soliton modes in this situation, we introduce some of distinct examples obtained from numerical analysis (Type I, II, and III summarized in Fig.~\ref{fig_soliton_types} and Table~\ref{phase_table}). 

In Type I, we combined a connected/non-repulsive phase with $\Theta_1=\{a_i=1.5, r_i=1,\bar\theta_i=2\,\vert\, i=1,\cdots10\}$ and a disconnected/non-repulsive phase with $\Theta_2=\{a_i=0.5, r_i=1,\bar\theta_i=2\,\vert\, i=1,\cdots10\}$ with the middle spacing of $a=2$. In this case, we observe a full transmission of a soliton from one end to the other even in the presence of the boundary between the two subsystems. The soliton keeps moving back and forth.
In Type II, we combined a disconnected/non-repulsive phase with $\Theta_1=\{a_i=1, r_i=2,\bar\theta_i=2\,\vert\, i=1,\cdots10\}$ and another disconnected/non-repulsive phase with $\Theta_2=\{a_i=0.5, r_i=1,\bar\theta_i=2\,\vert\, i=1,\cdots10\}$ with $a=2$, where both of them are in the non-repulsive state. In this case, we observe a complete reflection of a soliton at the boundary between the subsystems. As the soliton approaches to the boundary from the right subsystem, the rotors near the boundary in the left subsystem go up and down, and the soliton is totally reflected back into the right subsystem. We observe that the soliton goes back and forth only within the right subsystem. In Type III, we combined a disconnected/non-repulsive phase with $\Theta_1=\{a_i=1,r_i=2,\bar\theta_i=2\,\vert\,i=1,\cdots,10\}$ and connected/repulsive phase with  $\Theta_2=\{a_i=0.5,r_i=1,\bar\theta_i=0.2\,\vert\,i=1,\cdots,10\}$ with $a=2$. In this case, we also observe that the soliton is totally reflected at the boundary similarly to Type II, so its motion is confined in only the right subsystem. The rotors near the right end of the left subsystem stay lifted after the first reflection of the soliton at the boundary in the middle, and they go back to the initial positions after the second reflection.

\section{Discussion}\label{sec5}
We have constructed new soliton solutions in hybrid systems by connecting two linkages with distinct parameter sets.
Even though an explicit form of the solution is not available (which seems impossible in general), the trajectory of this dynamics can be understood via the topology of configuration space.
It should be pointed out that it is not clear whether or not the continuum limit is appropriate in our discussion.
Although one would obtain a continuous equation in the large $N$ limit, it potentially loses rich features appearing in discrete systems. In our formulation, solitons in a one-dimensional linkage system are trajectories that are approximated by constraint conditions. As long as the difference in the number of degrees of freedom of the rotors and the number of constraints is $1$, there is alway one-dimensional trajectory in the configuration space. Depending on how parameters of the linkage are chosen, trajectories can become complicated. When a linkage is realized under a uniform set of parameters, such as the case of Ref.~\cite{chenPNAS14}, trajectories can be simple. Even in hybrid systems considered our analysis, such as Type I in our simulation, we can still observe relatively simple trajectories. However, a certain choice of parameter sets can complicate such trajectories. 
Systems with parameter sets corresponding to Type II and III already exhibit nontrivial trajectories; a soliton is reflected at the boundary of two distinct subsystems.  
Combining two systems appears to produce further varieties of soliton solutions.

A soliton solution in the chain of four-bar linkages is reminiscent of the Fermi-Pasta-Ulam (FPU) problem, where a long-time dynamics reveals a recurrence phenomenon. (Concerning the FPU problem, see a modern account in Ref.~\cite{FPU}.)
The linkage discussed in our work can be regarded as a variant of the FPU problem for one-dimensional chains in the presence of nonlinear interactions.
In the case of four-bar linkages, a strong nonlinear interaction is given by an approximation of constraints by Eq.~(\ref{Eq:potential}).
This method can be investigated further.
The four-bar linkage has a relatively simple topology, which helps analysis of the dynamics, but the other type of linkages can be also studied in the same way.
Configurations spaces of linkages have been extensively investigated in robotics (Ref.~\cite{Farber}).

Even in the case of a four-bar linkage, one may encounter the critical set of parameters belonging neither to connected nor disconnected phases. Full analysis including this critical case will lead to a further understanding of this strongly coupled system. An extension of our formulation to two-dimensional problem will be an interesting question.

\acknowledgements

The authors would like to thank Ryota Nakai for discussions which initiated this project, Koji Hasegawa for helpful conversations on history and backgrounds of integrable systems, Natsuhiko Yoshinaga, Kei Yamamoto, and Mikio Furuta for fruitful discussions, Shin Hayashi for his careful reading and helpful comments on an early version of the paper, and
Motoko Kotani, the director of AIMR Tohoku University, for providing us the opportunity to work on this research project.


%

\end{document}